# Characteristics of the open star cluster Kronberger 60 using Gaia DR2

Tadross, A. L. *(altadross@gmail.com)*
and Hendy, Y. H. *(y.h.m.hendy@gmail.com)*

*National Research Institute of Astronomy and Geophysics, 11421 - Helwan, Cairo, Egypt*

**Abstract.** The estimation of the main parameters of star clusters is a very important target in astrophysical studies. Many of the open clusters listed in known catalogs have insufficient astrometric parameters. Improving the observations and analysis methods proved that some of them are asterisms and not true clusters. The most important aspect of using the Gaia DR2 survey lies in the positions, parallax, and proper motions for the cluster stars with the homogeneous photometry, which makes the membership candidates precisely determined. On this respect, several astor-photometric parameters of the open star cluster Kronberger 60 have been estimated. On studying the radial density profile, the radius is found to be 10.0±0.5 arcmin. It is located at a distance of 1935±90 pc, with an age of 800±50 Myr. Also, the reddening, the luminosity & mass functions, and the total mass of the cluster have been estimated as well. Our study is showing a dynamical relaxation behavior of Kronberger 60.

## 1 Introduction

Star clusters have been recognized as helpful tools in the structure and evolution of the Milky Way Galaxy (cf. Gilmore et al. 2012; Moraux 2016). Our knowledge of star cluster parameters, e.g. distances, ages, and color excesses are needed to obtain reliable conclusions. The available open cluster catalogs, e.g. Dias et al. (2002, 2014) obtain all the required data to produce dependable studies for deriving standardized parameters of open clusters (cf. Kharchenko et al. 2013; Dias et al. 2014; Sampedro et al. 2017). The catalog of Kharchenko et al. (2013) confirmed the PPMXL catalog of Roeser et al. (2010), who extends the basic data for a large set of clusters derived in a uniformed manner. There is large processing as a result of some differences, which cause some biases of the deduced parameters. Netopil et al. (2015) compared the clusters' parameters in many catalogs, inclusive Kharchenko et al. (2013), and concluded that there are discrepancies in distances, color excesses, and ages.

In fact, there are about 160 globular clusters and about 3000 open clusters in known catalogs (Kharchenko et al. 2013). If we extrapolate the sun nearness to all the galactic disc, we may recognize about 100,000 clusters. Many of them are recently discovered and their parameters should be measured in a confirmed method as well. We believe that most of them are going to be detected on account of Gaia, (Castro 2020 & Cantant 2018). In this context, we can investigate many open clusters that listed in global cata- logs, which have insufficient astrophysical parameters yet. The Gaia DR2 was released on 25 April 2018 for 1.7 billion high accuracy sources in astro- metric five-parameter of coordinates, proper motions in right ascension and declination, and parallaxes ($\alpha$, $\delta$, $\mu_\alpha \cos\delta$, $\mu_\delta$, $\pi$). In addition, magnitudes in three photometric filters ($G$, $G_{BP}$, $G_{RP}$) for more than 1.3 billion sources (Gaia Collaboration 2018). Gaia Archive is available through the web page (http://www.cosmos.esa.int/gaia).

We selected the open star cluster Kronberger 60, which is defined as an irregularly shaped cluster in constellation Auriga; embedded in a reflection Galactic Nebulae GN 06:00:08, (Magakian 2003; and Kronberger et al. 2006). Kronberger 60 is not defined in WEBDA Navigation Catalog (https://webda.physics.muni.cz/navigation.html), but it has just a few parameters in the literature. The main parameters of Kronberger 60 listed in Dais Catalog are taken from (Tadross 2009), who studied it photometrically for the first time depending on 2MASS database. Its diameter=1.6 arcmin, age=794 Myr, E(B-V)=0.84 mag, the distance from the Sun=1960 pc, from the Galactic center=10.5 kpc, and from the Galactic disk=162 pc.

The current work is a part of our continued series whose aim is to get the most astrophysical parameters of newly, previous unstudied and/or poorly studied open clusters using the foremost new missions. Therefore, we re-estimate the most astrophysical properties of Kronberger 60 using the database of GAIA DR2 mission. This paper is coordinated as follows: The data reductions are presented in Section 2. The structural parameters (cluster center and cluster size) are depicted in Section 3. The color-magnitude diagram analysis is clarified in Section 4. Luminosity - mass functions and the dynamical status of the cluster are described in Section 5. Finally, the conclusions of our study are summarized in Section 6.

## 2  Data reductions

The coordinates of the studied cluster is taken from Dias et al. (2002) catalog; (*https://wilton.unifei.edu.br/ocdb/*); version 3.5, 2016. Using Gaia DR2 (Gaia Collaboration 2018), it turned out that Gaia's data actually showed that clusters can be found larger than previously known. Therefore, the source data were downloaded from Gaia DR2 database service, located at the object's center with a radius of 40 arcmin to cover all the cluster's region, and contains all the data we need. The vector point diagram (VPD) of the proper motion in right ascension (pmRA) and declination (pmDE) is plotted as shown in Fig. 1.

The highest concentrated area is selected as a subset of the co-moving stars for which the current study stands on. It defines one of the membership criteria, i.e. the stars who move together in the same direction through space concerning the background field, Tadross (2018). The co-moving plot of Kronberger 60 is shown in Fig. 2. Stars are deemed cluster's members if its $3\sigma$ parallax error lies within the cluster mean parallax, and having the same speed and direction in the sky with respect to the background field ones, and, of course, located within the cluster's area (estimated from its radial density profile; see section 3.2). In addition, the membership of the cluster stars can be defined according to the cluster parallax's range that taken within errors ≤ 0.2 mas, and proper motions' errors in right ascension & declination ≤ 0.4 mas/yr. On the other hand, the relation between the magnitudes and the parallax shows that the distance range is narrow for the cluster members, while it is large for the field stars, as shown in Fig. 3. Also, the selected members form a specific track on the CMD (color-magnitude diagram) that represents the cluster's evolution curve, as illustrated in section 4.

## 3 Structural parameters

### 3.1 Cluster center

The cluster center is located at the most stellar density of the cluster's stars in its area. To determine the cluster center, a histogram of star counts is sketched in RA and DE using the database of Gaia DR2. To estimate the maximum central density of the cluster, we divided the studied area into equal sized bins in RA & DE and counted the stars in both directions. Fitting Gaussian profiles are applied to both directions RA and DE. The obtained central coordinates of Kronberger 60 is found in good agreement with the value taken from Simbad and Dias (2002) Catalogs, see Table 1.

### 3.2 Clusters' sizes

To obtain the radial density profile (RDP) of an open cluster, the surface stellar density should be derived by counting stars in concentric rings (shells) around the cluster's center. Dividing the numbers of stars by their corresponding areas. Then, we can get the calculated density in each shell by using the formula $R_i = N_i/A_i$, where $N_i$ is the number of stars within $i^{th}$ shell, and $A_i$ is the area of that shell. The density error in each shell is assumed to follow Poisson noise ($1/N^{1/2}$), where $N$ is the number of the stars counted in each shell. RDP for Kronberger 60 is shown in Fig. 4. The density distribution shows a peak near the cluster center and then decreases and flats after a certain point where the cluster density vanished into the field stars. The cluster limited radius, $R_{lim}$, can be estimated at that radius, which covers the whole cluster area and reaching sufficient constancy (cf. Tadross 2005). The radius $R_{lim}$ can be also estimated using the mathematic equation $R_{lim} = R_c ((f_0/3\sigma_{bg}) - 1)^{1/2}$ of Bukowiecki (2011). It is comparing $f(R)$ to the background density level, $f_b$, defined as $f_b = f_{bg} + 3\sigma_{bg}$, where $\sigma_{bg}$ is the uncertainty of $f_{bg}$. The empirical King's model (1966) is applied to the cluster as:

$$f(R) = f_{bg} + \frac{f_0}{1+(R/R_c)^2}$$

where $f_{bg}$, $f_0$ and $R_c$ are background, central density, and the core radius respectively. The core radius is the distance at which the stellar density equals half the central density. By fitting the King model to the radial density profile, the limited radius is found to be $R_{lim}$=10.0±0.5 arcmin, and the core radius $R_c$=0.38 arcmin.

The tidal radius of an open cluster is the distance from the cluster core at which the gravitational impact of the Galaxy equivalents to that of the cluster core. Calculating the overall mass of the studied cluster (see Sec. 5.2), the tidal radius can be calculated by applying the equation of Jeffries et al. (2001), $R_t = 1.46(M_c)^{1/3}$, where $R_t$ and $M_c$ are the tidal radius in parsecs and the overall mass of the cluster in solar mass, respectively. It is found to be $R_t$=8.4 pc. Also, the concentration parameter, which is an important parameter that shows us how the cluster is prominent in the sky. It can be calculated using the equation $C=log(R_{lim}/R_c)$ of Peterson and King (1975). It is found to be $C$=1.5, which declares that Kronberger 60 is a prominent cluster among the background field stars. All the structural parameters of the cluster ($f_0$, $f_{bg}$, $R_c$, $R_t$, and $C$) are estimated here for the first time and listed in Table 1.

## 4 Color-Magnitude Diagram

The Color-Magnitude diagram is a very important diagram for estimating the fundamental parameters of the candidate cluster. Age, distance modulus, color excess, and metallicity can be simultaneously determined by applying the theoretical stellar isochrones to the observed CMD and obtaining the visual best fit. It is substantial to recognize the sequence of the cluster from field stars because the stars in the cluster region are usually contaminated by the field ones.

The theoretical stellar isochrones according to Gaia data were obtained from the CMD 3.0 form of different ages (*http://stev.oapd.inaf.it/cgi−bin/cmd* 3.0) of Marigo et al. (2017). We used the recent solar metallicity value because of most open cluster studies based on isochrones fitting of nowadays solar metallicity $Z$=0.0152 (Caffau et al. 2009, 2011 & Steffi et al. 2018). The visual best fit of Kronberger 60 is verified with the isochrones of age=800 Myr, distance modulus m-M=12.05 mag (1935 pc), and color excess E(BP- RP)=0.30 mag, as shown in Fig. 5. On the other hand, the Gaussian fitting to the members yields a parallax mean value of 0.48 mas and distance of ≈2100 pc, which is in good agreement with our fitting. The black dots refer to the member stars lied within the cluster diameter, and located in the ranging parallax, within parallax's errors ≤ 0.2 mas, and proper motions' errors in RA & DE ≤ 0.4 mas/yr. Correspondingly, the distance from the galactic center $R_g$, the distances $X_\odot$, $Y_\odot$ from the Sun on the galactic plane, and the distance from the galactic plane $Z_\odot$ are estimated for Kronberger 60 and listed in Table 1. For more details about the calculations, see Tadross (2011).

We used the absorption ratio of the photometric systems in different wavelengths to the visual absorption ($A_\lambda/A_v$); Cardelli et al. (1989), O´Donnell (1994), and the Padova website CMD 3.0. The absorption ratios of Gaia DR2 are $A_G/A_v$=0.861, $A_{G_{BP}}/A_v$=1.062, and $A_{G_{RP}}/A_v$=0.651. These ratios have been used for correction of the magnitudes for the interstellar reddening and converting the color excess to E(B-V), where $R_v = A_v/E(B-V)=3.1$; E(B-V)=0.785 E(BP-RP).

## 5 Luminosity & mass functions, and dynamical state

An open star cluster exemplifies hundreds of stars having the same age and chemical structures but different masses. The luminosity and mass functions (LF & MF) of a cluster are counted on the size of the membership of that cluster. The main problem for studying the LF and MF is to exclude the field star contamination from the cluster's area. Using the RDP, the parallax, and proper motion conditions of Gaia DR2, we can separate the cluster members from the field stars in an acceptable way.

### 5.1 Luminosity function

Applying all the astrometric and photometric criteria mentioned above; counting the stars within the main sequence envelope in terms of the absolute magnitude $M_G$ that derived from the distance modulus of the cluster. We can draw a histogram of LF for Kronberger 60 as shown in the left-hand panel of Fig. 6. The luminosities and masses of the clusters' members can be given by applying the data of the isochrones of Marigo et al. (2017) at the age of the cluster. The magnitude bin intervals are selected to be suitable for stars in every bin for the most effective potential statistics of the LF and MF. Generally, we can infer that more massive stars are more centrally concentrated whereas the peak value lies at the fainter magnitude bin. The total luminosity of Kronberger 60 is found to be -3.0 mag. The lower panel of Fig. 6 represents the accumulated frequency distribution for Kronberger 60 as a function of radial distances and magnitudes. The mass segregation is visible as the higher masses (brighter stars) are expected to be settled toward the cluster center, while the lower masses (fainter stars) are residing in the outer region of the cluster.

## 5.2 Mass function

The LF and MF are inseparable according to the familiar relation called the mass-luminosity relation. Since we haven't an empirical transformation for such relation here, so we rely on the applied isochrones' data. The cluster members were divided into mass intervals of the absolute magnitude bins. The resulting MF of Kronberger 60 is shown in the right-hand panel of Fig. 6, whereas the red line refers to the initial mass function IMF- slope, which can be obtained from the subsequent equation:

$$\frac{dN}{dM} \propto M^{\alpha}$$

where $\frac{dN}{dM}$ is the number of stars in the mass interval $[M : (M + dM)]$ and $\alpha$ is the slope of the relation (where $\alpha = -2.35 \pm 0.24$), which indicates that the estimated masses of the studied cluster lie in the range of Salpeter (1955). The slope of that relation according to Kronberger 60 is found to be $\alpha = -2.26$, which is in good agreement with Salpeter's one. The total mass of the target cluster is calculated by integrating the masses of all the cluster members (Sharma et al. 2006). All faint stars below G=21 mag (or lower than 0.4 $M_{\odot}$) have been excluded (Lamers et al. 2005). The total mass of Kronberger 60 is found to be 190 $M_{\odot}$.

## 5.3 Dynamical state of the clusters

The relaxation time is the time scale in which a cluster loses all traces of its initial conditions and represented by $T_R$. It is the characteristic time-scale for a cluster to reach the equivalent level of energy and given by Spitzer & Hart (1971) as:

$$T_R = \frac{8.9 \times 10^5 \sqrt{N} \times R_h^{1.5}}{\sqrt{\langle m \rangle} \times log(0.4N)}$$

where $N$ is the number of the cluster members, $R_h$ is the radius containing half of the cluster mass in parsecs, and $\langle m \rangle$ is the average mass of the cluster in the solar unit, assuming that $R_h$ equals half of the cluster radius. Then, the relaxation time is calculated for Kronberger 60 and found to be $T_R=34$ $Myr$. The comparison of the age of the cluster (800 $Myr$) with its relaxation time indicates that the relaxation time is much smaller than the age of the cluster. From the dynamical evolution parameter $\tau = Age/T_R$, we conclude that Kronberger 60 is dynamically relaxed, where its age is 23.5 times older than the relaxation time.

# 6  Conclusions

We presented here deeper analysis for refining and estimating the fundamental parameters of the poor studied open cluster Kronberger 60, which is an irregularly shaped cluster in constellation Auriga embedded in a reflection Nebulae. The main parameters of Kronberger 60, which are obtained in the literature are taken from (Tadross 2009) who studied it photometrically for the first time relied on 2MASS database. It is obvious that, using Gaia DR2 database, we obtained more precise parameters but larger size than previously known. The main findings of our new analysis are listed in Table 1. The new and old results can be compared.

**Acknowledgments**

This work is a part of a project called "IMHOTEP" program No. 42088ZK between Egypt and France. We are grateful to the Academy of Scientific Research and to the Service of scientific cooperation in both countries for giving us the opportunity to work through this project and pro- vide the necessary support to accomplish our scientific goals. We thank Prof. David Valls-Gabaud, our companion in the project, for his valuable discussion during our visit to Paris in Sept. 2019. This work has made use of data from the European Space Agency (ESA) mission Gaia processed by the Gaia Data Processing and Analysis Consortium (DPAC),
(*https*: *//www.cosmos.esa.int/web/gaia/dpac/consortium*).
Funding for the DPAC has been provided by national institutions, in particular, the institutions participating in the Gaia Multilateral Agreement.

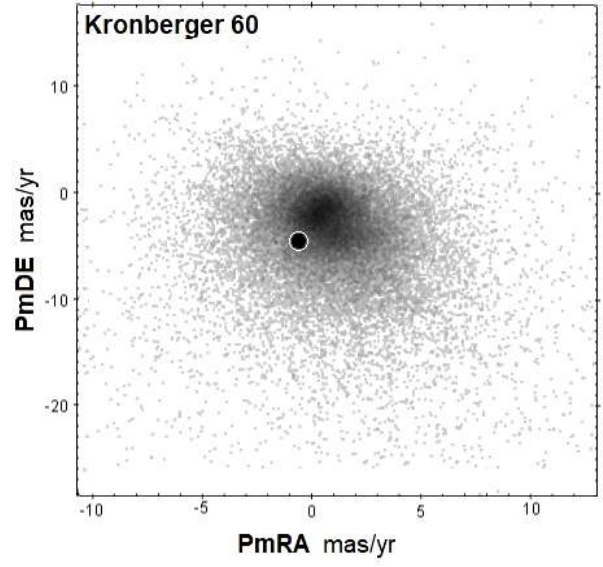

Figure 1: The vector point diagram *VPD* of Kronberger 60, from which the highest concentrated area is selected for the current study.

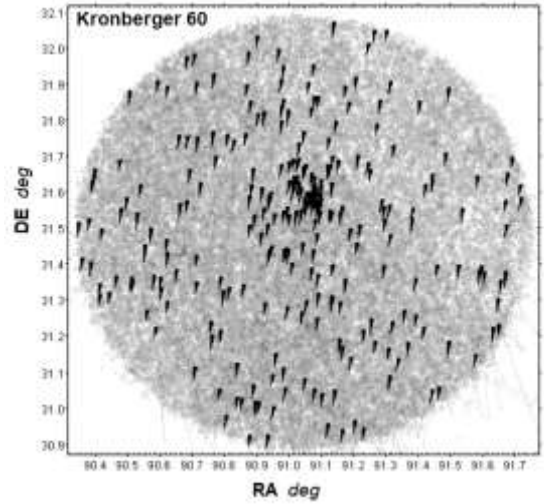

Figure 2: The co-moving stars of Kronberger 60. It is selected as a subset from the VPD of the cluster and defines the membership criteria, i.e. the stars who travel together in the same direction through space and lay in the limited size of the cluster.

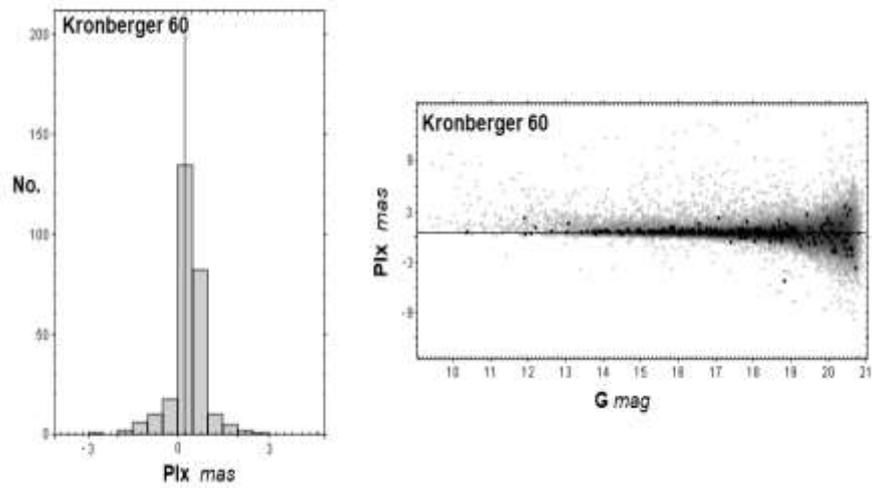

Figure 3: The left-hand panel shows the parallax range of Kronberger 60, while the right-hand panel presents the relation between the magnitude and the parallax of Kronberger 60. It shows that the distance range is narrow for the cluster members, while it is large for the field ones. The solid lines represent the mean value of the cluster's parallax (0.48 mas).

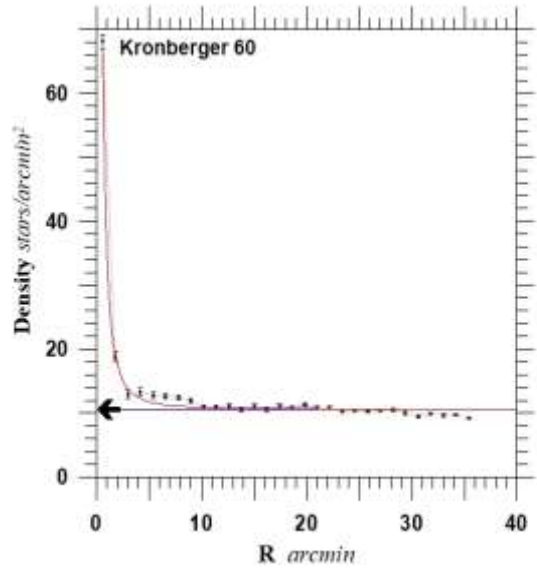

Figure 4: Surface density profile of Kronberger 60. Error bars are determined from sampling statistics ($=1/N^{1/2}$ where $N$ is the number of stars used in the density estimation at that point). The smooth red line represents the fitted profile of King (1966). The limited radius is found to be 10.0±0.5 arcmin. The arrow indicates the estimated background field of the cluster.

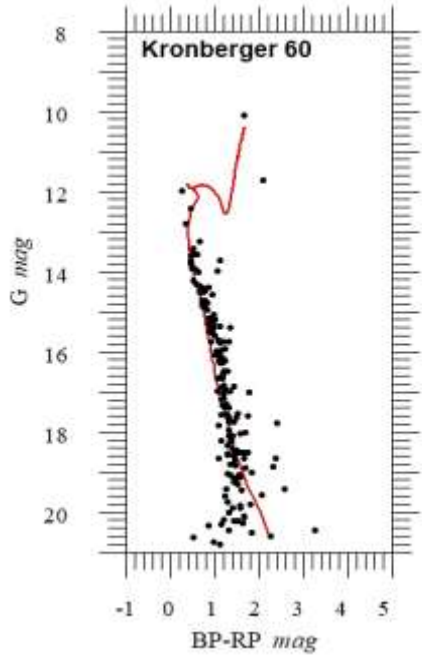

Figure 5: The fitted CMD of Kronberger 60. The theoretical isochrones of solar metallicity $Z=0.0152$ with the age of 800 Myr of Marigo et al. (2017) has been applied. The visual best fit is taken at distance modulus m-M=12.05 mag, and color excess E(BP-RP)=0.30 mag. The black dots refer to the members lied within the cluster area, and located in the ranging parallax, within parallax errors ≤ 0.2 mas, and proper motions' errors in RA & DE ≤ 0.4 mas.

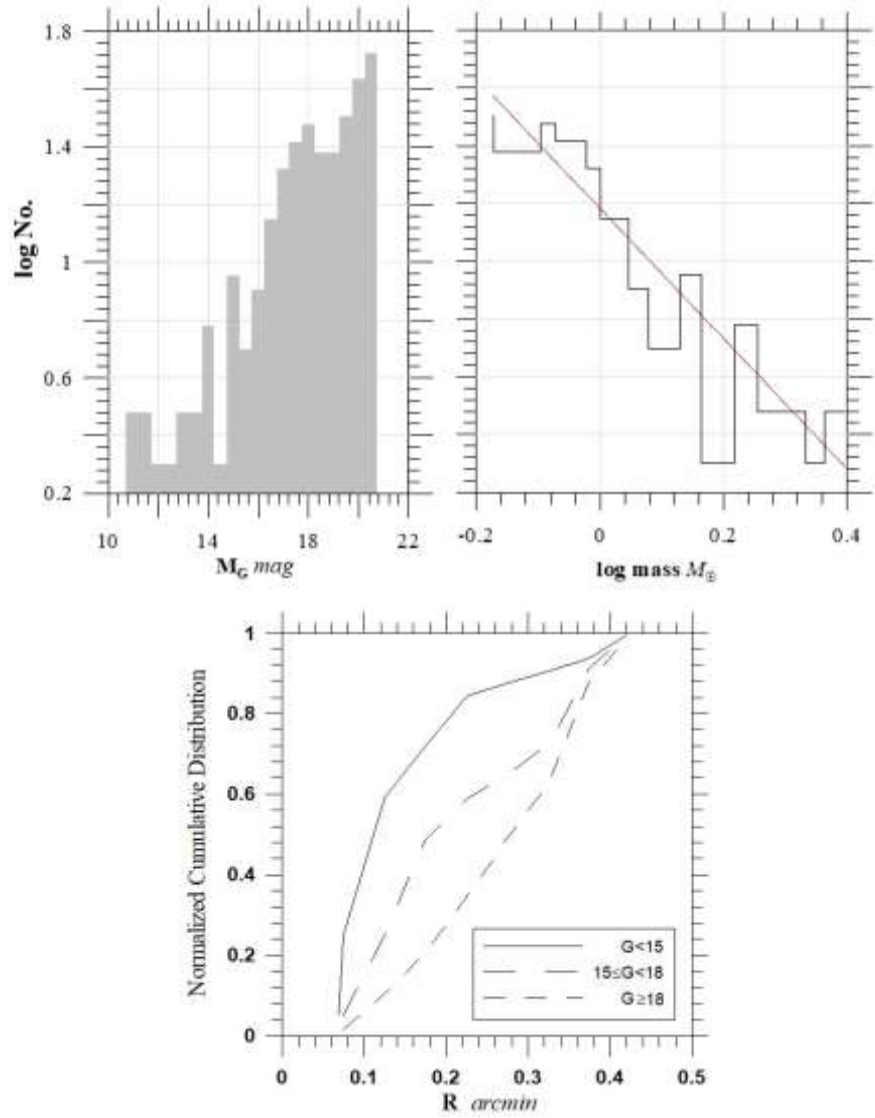

Figure 6: The luminosity & mass functions, and mass segregation of Kronberger 60. In the left-hand panel, the gray histogram represents the luminosity distribution of the cluster. The right-hand panel represents the mass distribution of the cluster, the red line shows the linear fitting of the initial mass function *IMF*-slope, which is -2.26. The lower panel represents the ac- cumulated frequency distribution for Kronberger 60 as a function of distance from the cluster center and magnitudes.

Table 1: The astrophysical parameters of Kronberger 60.

| Parameter | New/ (Old) values |
|---|---|
| $\alpha$(2000)  deg | 91.0318 / (91.0420) |
| $\delta$(2000)  deg | 31.5360 / (31.4956) |
| $l$(2000)  deg | 179.7676 / (179.8074) |
| $b$(2000)  deg | 4.7620 / (4.7500) |
| Age  $Myr$ | 800/ (794) |
| m-M  $mag$ | 12.05/ (12.20) |
| E(BP-RP)  $mag$ | 0.30/ (—) |
| E(B-V)  $mag$ | 0.23/ (0.84) |
| dist.  $pc$ | 1935/ (1960) |
| $R_g$  $kpc$ | 10.27/ (10.50) |
| $X_\odot$  $pc$ | 1926/ (1953) |
| $Y_\odot$  $pc$ | 6.50/ (6.60) |
| $Z_\odot$  $pc$ | 160/ (162) |
| $R_{lim}$  arcmin | 10.0/ (0.80) |
| $f_0$  stars/arcmin$^2$ | 199/ (—) |
| $f_{bg}$  stars/arcmin$^2$ | 10.6/ (—) |
| $R_c$  arcmin | 0.38/ (—) |
| $R_t$  $pc$ | 8.4/ (—) |
| C | 1.5/ (—) |
| Mem.  stars | 180/ (—) |
| $\alpha$ | -2.26/ (—) |
| Total lumin.  $mag$ | -3.0/ (—) |
| Total mass  $M_\odot$ | 190/ (—) |
| Relax. time  $Myr$ | 34/ (—) |
| $\tau$ | 23.5/ (—) |

$R_g$ is the distance from the galactic center. $X_\odot$, $Y_\odot$, and $Z_\odot$ are the Galactocentric coordinates of the cluster. The Y-axis connects the Sun to the Galactic Center, while the X-axis is perpendicular to Y-axis. $Y_\odot$ is positive towards the Galactic anti-center, and $X_\odot$ is positive in the first and second Galactic quadrants (Lynga 1982). *Mem* is the number of the cluster members, $\alpha$ is the slope of the mass function, and $\tau$ is the dynamical evolution parameter. $R_{lim}, f_0, f_{bg}, R_c, R_t$, and $C$ are the structural parameters of the cluster. For comparisons, the values between brackets refer to that results obtained by Tadross (2009).